\documentclass[twoside,twocolumn,9pt]{article}
\usepackage{extsizes}
\usepackage[super,sort&compress,comma]{natbib} 
\usepackage[version=3]{mhchem}
\usepackage[left=1.5cm, right=1.5cm, top=1.785cm, bottom=2.0cm]{geometry}
\usepackage{balance}
\usepackage{times,mathptmx}
\usepackage{sectsty}
\usepackage{graphicx} 
\usepackage{lastpage}
\usepackage[format=plain,justification=raggedright,singlelinecheck=false,font={stretch=1.125,small,sf},labelfont=bf,labelsep=space]{caption}
\usepackage{float}
\usepackage{fancyhdr}
\usepackage{fnpos}
\usepackage[english]{babel}
\usepackage{array}
\usepackage{droidsans}
\usepackage{charter}
\usepackage[T1]{fontenc}
\usepackage[usenames,dvipsnames]{xcolor}
\usepackage{setspace}
\usepackage[compact]{titlesec}

\usepackage{amsmath}
\usepackage{hyperref}

\graphicspath{{./figures/}}

\newcommand{\AB}{NH$_3$BH$_3$}
\newcommand{\ABx}{NH$_3$BH$_2\mathcal{X}$}
\newcommand{\NH}{NH$_3$}
\newcommand{\BH}{BH$_3$}

\definecolor{cream}{RGB}{222,217,201}

\begin{document}

\pagestyle{fancy}
\thispagestyle{plain}
\fancypagestyle{plain}{

}

\makeFNbottom
\makeatletter
\renewcommand\LARGE{\@setfontsize\LARGE{15pt}{17}}
\renewcommand\Large{\@setfontsize\Large{12pt}{14}}
\renewcommand\large{\@setfontsize\large{10pt}{12}}
\renewcommand\footnotesize{\@setfontsize\footnotesize{7pt}{10}}
\makeatother

\renewcommand{\thefootnote}{\fnsymbol{footnote}}
\renewcommand\footnoterule{\vspace*{1pt}%
\color{cream}\hrule width 3.5in height 0.4pt \color{black}\vspace*{5pt}} 
\setcounter{secnumdepth}{5}

\makeatletter 
\renewcommand\@biblabel[1]{#1}            
\renewcommand\@makefntext[1]%
{\noindent\makebox[0pt][r]{\@thefnmark\,}#1}
\makeatother 
\renewcommand{\figurename}{\small{Fig.}~}
\sectionfont{\sffamily\Large}
\subsectionfont{\normalsize}
\subsubsectionfont{\bf}
\setstretch{1.125} 
\setlength{\skip\footins}{0.8cm}
\setlength{\footnotesep}{0.25cm}
\setlength{\jot}{10pt}
\titlespacing*{\section}{0pt}{4pt}{4pt}
\titlespacing*{\subsection}{0pt}{15pt}{1pt}

\fancyfoot{}
\fancyfoot[RO]{\footnotesize{\sffamily{1--\pageref{LastPage} ~\textbar  \hspace{2pt}\thepage}}}
\fancyfoot[LE]{\footnotesize{\sffamily{\thepage~\textbar \hspace{2pt} 1--\pageref{LastPage}}}}
\fancyhead{}
\renewcommand{\headrulewidth}{0pt} 
\renewcommand{\footrulewidth}{0pt}
\setlength{\arrayrulewidth}{1pt}
\setlength{\columnsep}{6.5mm}
\setlength\bibsep{1pt}

\makeatletter 
\newlength{\figrulesep} 
\setlength{\figrulesep}{0.5\textfloatsep} 

\newcommand{\topfigrule}{\vspace*{-1pt}%
\noindent{\color{cream}\rule[-\figrulesep]{\columnwidth}{1.5pt}} }

\newcommand{\botfigrule}{\vspace*{-2pt}%
\noindent{\color{cream}\rule[\figrulesep]{\columnwidth}{1.5pt}} }

\newcommand{\dblfigrule}{\vspace*{-1pt}%
\noindent{\color{cream}\rule[-\figrulesep]{\textwidth}{1.5pt}} }

\makeatother

\twocolumn[
  \begin{@twocolumnfalse}
\vspace{3cm}
\sffamily
\begin{tabular}{m{4.5cm} p{13.5cm} }

 & \noindent\LARGE{\textbf{Lowering the hydrogen desorption temperature of \AB\ through B-group substitutions}} \\
\vspace{0.3cm} & \vspace{0.3cm} \\

 & \noindent\large{\textbf{Evan Welchman and Timo Thonhauser\textit{$^{a}$}}}\\

\vspace{0.5cm}

\vspace{0.4cm} & \noindent\normalsize{
We present \emph{ab initio} results for substitutions in the promising
hydrogen-storage material \AB\ to lower its hydrogen desorption
temperature. Substitutions have already been investigated with
significant success recently, but in all cases a \emph{less}
electronegative element is substituted for the \emph{protic} hydrogen in
the \NH\ group of \AB. We propose a different route, substituting the
\emph{hydridic} hydrogen in the BH$_3$ group with a \emph{more}
electronegative element. To keep the gravimetric density high, we focus
on the second period elements C, N, O, and F, all with higher
electronegativity compared to H. In addition, we investigate Cu and S as
possible substitutions. Our main results include Bader charge analyses,
hydrogen binding energies, and kinetic barriers for the hydrogen release
reaction in the gas phase as well as in the solid.  While the different
substituents show varying effects on the kinetic barrier and thus
desorption temperature---some overshoot our goal while others have
little effect---we identify
Cu as a very promising substituent, which lowers the reaction barrier by
approximately 37\% compared to \AB\ and thus significantly influences the hydrogen
desorption temperature.} \\

\end{tabular}

 \end{@twocolumnfalse} \vspace{0.6cm}

  ]

\renewcommand*\rmdefault{bch}\normalfont\upshape
\rmfamily
\section*{}
\vspace{-1cm}


\footnotetext{Department of Physics, Wake Forest University, Winston-Salem, NC 27109, USA.\ }
\footnotetext{\textit{$^{a}$~thonhauser@wfu.edu }}



\section{Introduction}\label{sec:introduction}

One of the greatest technical challenges facing society is the need to end our
dependence on fossil fuels. Shifting to hydrogen as an alternative
energy carrier presents an ideal solution, but is not yet practical, as
its storage is one of the remaining bottlenecks in the
technology.\cite{Zuttel_2004:hydrogen_storage,
Basic_Research_Needs_2004} While conventional pressurized tanks cannot
safely store hydrogen at sufficient density to be used as an equivalent
substitute for gasoline in mobile applications, incorporating the
hydrogen into the molecular structure of a material might reach the
required density target set by the DOE.\cite{DOE_Targets_Onboard_2009, Harrison_2014:materials_hydrogen, Graetz_2009:new_approaches} Amongst
such materials, ammonia borane (\AB\ or AB) has been the subject of
myriad studies and is one of the most promising hydrogen storage
materials with a gravimetric storage density of
19.6~mass\%.\cite{Xiong_2008:high-capacity_hydrogen,
Chua_2011:development_amidoboranes, Swinnen_2010:potential_hydrogen,
Hamilton_2009:b-n_compounds, Heldebrant_2008:effects_chemical,
Marder_2007:will_we} 

Unfortunately, besides its favorable gravimetric hydrogen storage
density, \AB\ starts to release its hydrogen only at elevated
temperatures (12~mass\% above $140^\circ$C, corresponding to the release
of one H$_2$ per \AB\ unit).\cite{Stowe_2007:situ_solid,Kobayashi_2014:mechanism_solid-state}
One of the most notable advances in this area was the substitution of a
protic hydrogen in the NH$_3$ group with a Group I metal to form an
alkali-metal amidoborane ($\mathcal{M}$NH$_2$BH$_3$ where
$\mathcal{M}$~=~Li, Na).\cite{Xiong_2008:high-capacity_hydrogen} These
materials demonstrate hydrogen desorption temperatures significantly
lower than that of \AB---90$^\circ$C with 10.9~mass\% and 8.9~mass\% for
Li and Na, respectively. However, a further lowering is desirable, as
for instance, the maximum operating temperature for a proton-exchange
membrane fuel cell is 80$^\circ$C.\cite{Yang_2010:high_capacity}

To this end, we propose a similar, albeit different, modification to the
material's molecular structure. Rather than substitute a metal for a
protic hydrogen in the \NH\ group, we instead replace a hydridic
hydrogen in the \BH\ group with a more electron-accepting element. To
keep the gravimetric density high, our obvious choices are the second
period elements with the corresponding electronegativities C (2.55), N
(3.05), O~(3.44), and F~(3.98), all higher than the value of 2.20 for
H.\cite{Allred_1961:electronegativity_values} We also consider using S
(2.58) and Cu (1.90) as substituents. Cu is included because its single
4$s$ electron (but otherwise closed shells) makes it electronically
similar to H. The electronegativity of Cu is actually lower than that
of H, but we find that in the monomer, Cu still prefers to bind to the \BH\ group rather
than the \NH\ group by 0.19~eV. We include S for
reasons that will be elaborated on later.  Note that these kinds of
substitutions have not yet been tried in the hydridic group, as their
synthesis is more complicated.\footnote{W. Mao, private communication.}
Also note that substitutions of the protic hydrogen in the \NH\ group
are more straightforward in principle and ``less intrusive'' to the
electronic structure, as all alkali metals have a single $s$ electron in
their outer shell and H has already the highest electronegativity of all
alkali metals. On the other hand, our proposed substitutions of the
hydridic hydrogen in the \BH\ group are likely to create radicals and
constitute an unconventional and major rearrangement of the bonding and
charge density in the entire molecule. Nonetheless, according to our
calculations, all substitutions create stable molecules with a negative
cohesion energy and no imaginary modes in their vibrational spectra,
indicating that their synthesis is in principle possible.

Several studies have aimed to investigate the hydrogen desorption
mechanism for both \AB\ and the alkali-metal amidoboranes using
experimental, density functional theory, and quantum chemistry
methods.\cite{Shevlin_2011:dehydrogenation_mechanisms,
Stowe_2007:situ_solid, Zhong_2012:first-principles_investigation,
Miranda_2007:ab_initio, Nguyen_2007:computational_study, Wolstenholme_2012:thermal_desorption,
Choi_2014:kinetics_study} The computational parts of these studies have
focused mostly on the interactions of molecules in the gas phase. We aim
here to take full advantage of density functional theory in a plane-wave
implementation and study not only gas-phase effects, but also the
release and formation of molecular hydrogen in the solid state, taking
into account the effects of nearby molecules on the reaction. To this
end, this manuscript is organized into two main sections: First, we
examine only gas-phase structures in order to validate our methods as
well as filter out substituents from our list that yield less desirable
results. Then, we perform calculations in the solid phase, again
filtering out substituents, and encourage the synthesis of the structures that show improved hydrogen desorption properties compared to \AB.

\section{Computational Details}\label{sec:comp_details}

We have performed \emph{ab initio} calculations at the density
functional theory (DFT) level, as implemented in the plane-wave code
\textsc{Vasp} (version 5.3.3).\cite{Kresse_1996:efficient_iterative,
Kresse_1999:ultrasoft_pseudopotentials} Since ammonia
borane is a strong van der Waals complex,\cite{Chen_2010:situ_x-ray,
Lin_2008:raman_spectroscopy} it is crucial to include van der Waals
forces in our simulations. Thus,
all calculations have been performed with the
vdW-DF1\cite{Dion_2004:van_waals, Thonhauser_2007:van_waals,
Langreth_2009:density_functional, Berland_2015:van_waals}
exchange-correlation functional (i.e.\ revPBE exchange and LDA
correlation in addition to the truly nonlocal correlation) with PAW
pseudopotentials and an energy cutoff of 500~eV. Calculations in the gas
phase were done in a 12~\AA\ cubic box at the $\Gamma$-point. All
structures were optimized until the forces on each atom were less than
1~meV/\AA. All calculations were performed spin-polarized, as some of
our substitutions result in unpaired electrons.

In the solid, structure searches have been carried out using the
Universal Structure Predictor: Evolutionary Xtallography (USPEX),
\cite{Oganov_2006:crystal_structure, Lyakhov_2013:new_developments, Zhu_2012:constrained_evolutionary} with
structural optimizations done in \textsc{Vasp}.  We performed structure
searches with two molecules per unit cell to correspond with the ground
state structure of pure AB. In addition to these randomly-generated
structures, we also included seed structures generated by modifying the
low-temperature phase of pure AB. Each search spanned five generations,
totaling over 150 structural optimizations per substituent. Because
lattice parameters were allowed to (and did) vary, k-point meshes
differed between structures, but their density was kept constant, i.e.\
the number of k-points $k_i$ in a given reciprocal space dimension was
given by $k_i = (0.1 \times l_i)^{-1}$ where $l_i$ is the length of a
lattice vector. Each of these structures was optimized until all ionic forces
dropped below 10~meV/\AA.

Nudged elastic band (NEB) calculations were also performed in
\textsc{Vasp}, utilizing the climbing image implementation from the \textsc{Vtst} Tools
package.\cite{Henkelman_2000:climbing_image,
Henkelman_2000:improved_tangent} For the solid phase, these calculations
were performed on $2\times2\times2$ supercells to minimize finite-size effects,
with either 5 or 7 intermediate images, each converged to forces of less than 50~meV/\AA.

At low temperatures (0 $\sim$ 225~K), \AB\
exhibits an orthorhombic structure with space group \emph{Pmn}2$_{1}$.
Heated above 225~K, it undergoes a phase transition to a body-centered
tetragonal structure with space group \emph{I}4\emph{mm}. In both cases
there are two \AB\ molecules per conventional unit cell and the main
difference is that the N--B backbone of the molecules are parallel in
the high-temperature phase, while they are slightly tilted with respect
to each other in the low-temperature phase.  Although we are mostly
interested in hydrogen desorption at room temperature or higher,
unfortunately, the high-temperature phase is not accessible for standard
transition state searches, as it tries to relax towards the
low-temperature phase. As such, all our calculations are done in the
low-temperature phase.

\section{H$_2$ desorption in the gas phase}\label{sec:gasphase}
\subsection{Structure analysis}

\begin{table}
\caption{\label{tab:gas_bind} Binding energies (eV) of atoms in gas
phase molecules \ABx\ with $\mathcal{X}$ = H, F, O, S, Cu. H itself is
not a substitution, resulting in the original \AB\ molecule, but is
given here for reference. Atom designations correspond to those in
Fig.~\ref{fig:AB_label} and substituent atoms always replace the H in
position B3.}
\begin{tabular*}{\columnwidth}{@{} l @{\extracolsep{\fill}} c c c c c r @{}}\hline
$\mathcal{X}$   & N1       & N2       & N3       & B1       & B2       & B3$^\text{sub}$\\\hline
H$^\text{ref}$  & $-4.140$ & $-4.140$ & $-4.140$ & $-4.592$ & $-4.592$ & $-4.592$       \\
             F  & $-4.137$ & $-4.138$ & $-4.138$ & $-4.609$ & $-4.609$ & $-6.962$       \\
             O  & $-4.296$ & $-4.297$ & $-4.297$ & $-1.058$ & $-1.058$ & $-6.120$       \\
             S  & $-3.924$ & $-4.175$ & $-4.175$ & $-1.982$ & $-1.982$ & $-4.336$       \\
             Cu & $-2.638$ & $-2.610$ & $-2.610$ & $-4.391$ & $-4.391$ & $-2.659$       \\[4ex]
\end{tabular*}
\end{table}

\begin{figure}\centering
 \includegraphics[width=0.45\columnwidth]{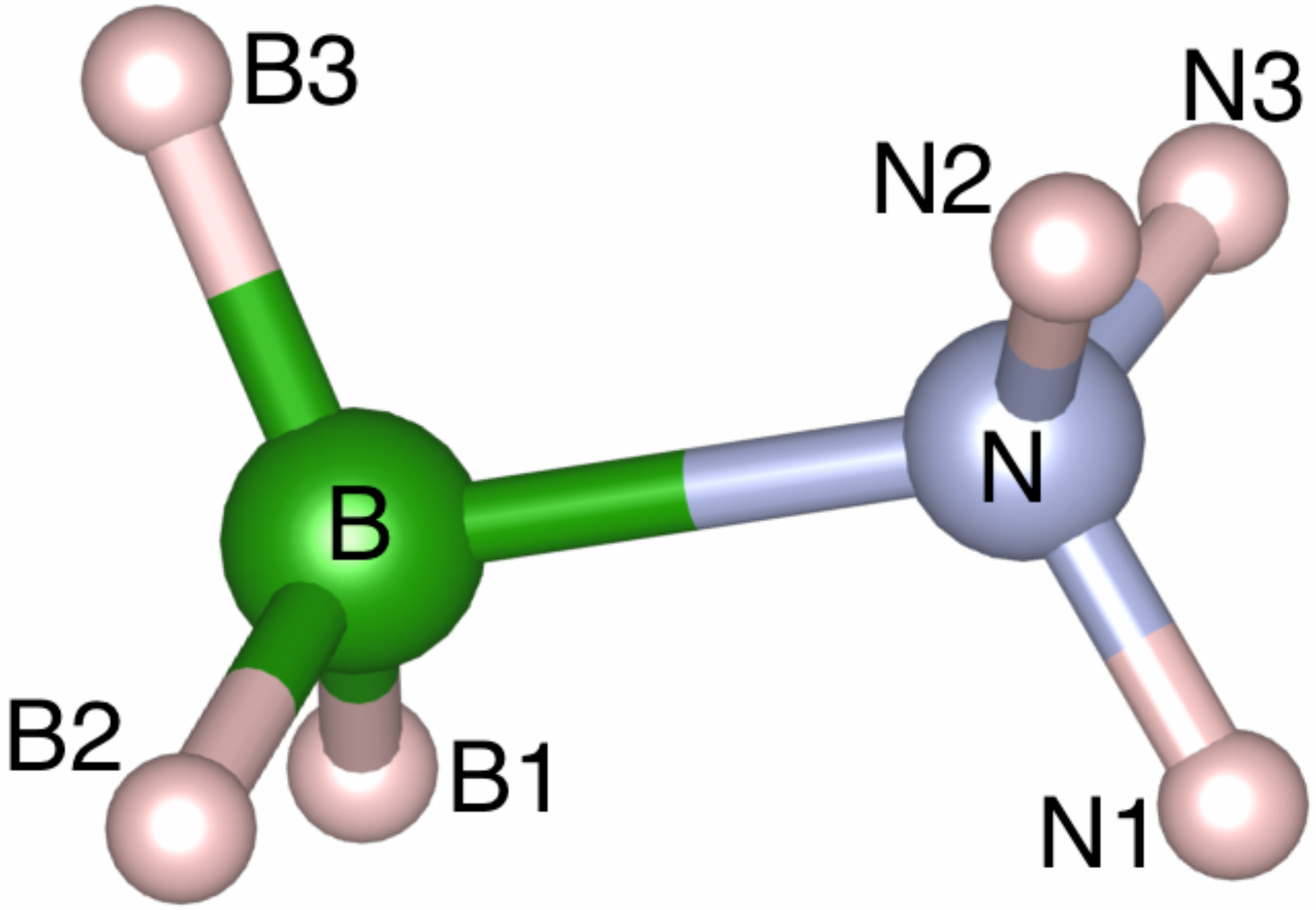}
 \caption{\label{fig:AB_label} Diagram of an \AB\ molecule, including
designations used to refer to each H atom. All substitutions have been
done at the B3 position.}
\end{figure}

The first step in examining these prospective materials resulting from
our proposed substitutions is optimizing the molecules's structures in
the gas phase. For the purposes of this investigation, we are only
interested in materials with hydrogen desorption properties similar to
those of AB, so we preferentially selected only the substituents that
formed NH$_3$BH$_2\mathcal{X}$ molecules, retaining AB's original
molecular shape, to continue study. We eliminated C and N from our
original candidate pool at this stage because simple structural
optimizations indicated that these substitutions did not allow for the
formation of the NH$_3$BH$_2\mathcal{X}$ geometry. Those substitutions,
due to their strong radical nature, resulted in a complete rearrangement
of atoms in the \BH\ group. Also note that for each of the substituents used, our
calculations indicate that substitution is energetically favorable in
the B-group relative to substitution in the N-group of an isolated
molecule.

We next found binding energies for each atom in gas phase molecules with
O, F, S, and Cu substitutions, given in Table~\ref{tab:gas_bind}. We
expect that a successful substitution would lower the strength with
which H atoms bind to the molecule, making this information a crude
indication of the magnitude of a substituent's effect on the molecule.
The F substitution appears to have little effect on the binding energies
compared to the H reference, while O, S, and Cu elicit significant
changes. O and S reduce the binding energies of H atoms on the B side of
the molecule whereas, interestingly, Cu affects the N side more
significantly.  

Although charge partitioning schemes are not unique, one can gain insight into the reasons for these changes in binding energies by looking at the Bader charge analysis (shown in Table \ref{tab:bader}) of
an isolated molecule in gas phase. As expected, in the pure AB case, B
donates three electrons to the H atoms bound to it and N takes the 3
electrons from the H atoms bound to it, but there is very little charge
transfer between the two ends of the molecule. In this picture, H atoms in the B group are nearly perfectly accepting one electron. F and S substitutions do little to change the picture, but O and Cu substitutions have stronger effects.

The O atom takes a greater share of the charge from the nearby H atoms, resulting in a large decrease in these H atoms' binding energies and a correspondingly large decrease in the gas phase H$_2$ desorption barrier (see section \ref{sec:gas} and Table \ref{tab:NEB_gas}). While this is the desired effect, it is actually overshooting to a point where the O-substituted molecule becomes unstable in the solid phase, releasing hydrogen near 0~K. We would expect that S would produce a similar, but less pronounced effect in terms of altering the molecule's electronic structure. The charge analysis and binding energies support this hypothesis, with S taking more charge than its neighbor H atoms and significantly lowering these H atoms' binding energies, although displaying a higher H$_2$ desorption barrier (again, see Table \ref{tab:NEB_gas}). Clearly, the overall effect of these substitutions cannot be fully explained with changes in the molecule's charge density or binding energies.

Where O and S caused a decrease in the binding energies of H atoms in
the B group, Cu substitution results in a decrease of binding energies
for H atoms in the N group. The Cu substitution's effect on the Bader
charge is similarly distinctive, with B only donating 2.47 electrons, of
which Cu takes only 0.22.

\begin{table}
\caption{\label{tab:bader} Bader charge analysis (in units of $e$) for
isolated gas phase molecules of \ABx. Numbers given demonstrate the
change in charge after the substitution is made at position B3 and
reflect the number of electrons gained relative to the neutral atom.}
\begin{tabular*}{\columnwidth}{@{} l @{\extracolsep{\fill}} cccccccr @{}}\hline
$\mathcal{X}$  & N1      & N2      & N3      & N    & B       & B1   & B2   & B3$^\text{sub}$\\\hline
H$^\text{ref}$ & $-1.00$ & $-1.00$ & $-1.00$ & 2.96 & $-3.00$ & 0.99 & 1.01 & 1.03\\
           F   & $-1.00$ & $-1.00$ & $-1.00$ & 2.97 & $-3.00$ & 1.04 & 1.03 & 0.97\\
           O   & $-1.00$ & $-1.00$ & $-1.00$ & 2.94 & $-3.00$ & 0.88 & 0.87 & 1.30\\
           S   & $-1.00$ & $-1.00$ & $-1.00$ & 3.00 & $-3.00$ & 0.98 & 0.98 & 1.03\\
           Cu  & $-1.00$ & $-1.00$ & $-1.00$ & 3.01 & $-2.47$ & 1.12 & 1.12 & 0.22\\
\end{tabular*}
\end{table}

\subsection{H$_2$ release barriers in the gas phase}\label{sec:gas}

\begin{table}
\caption{\label{tab:NEB_gas} Hydrogen desorption barriers (eV) from a
single gas-phase molecule of \ABx\ resulting in NH$_2$BH$\mathcal{X}$
and H$_2$.}
\begin{tabular*}{\columnwidth}{@{} l @{\extracolsep{\fill}} r @{}}\hline
$\mathcal{X}$ & gas phase H$_2$ desorption barrier [eV]\\\hline
H$^\text{ref}$ & 1.706 \\
             F & 1.675 \\
             O & 1.014 \\
             S & 1.812 \\
            Cu & 1.697 \\[4ex]
\end{tabular*}
\end{table}

\begin{figure}
\includegraphics[width=\columnwidth]{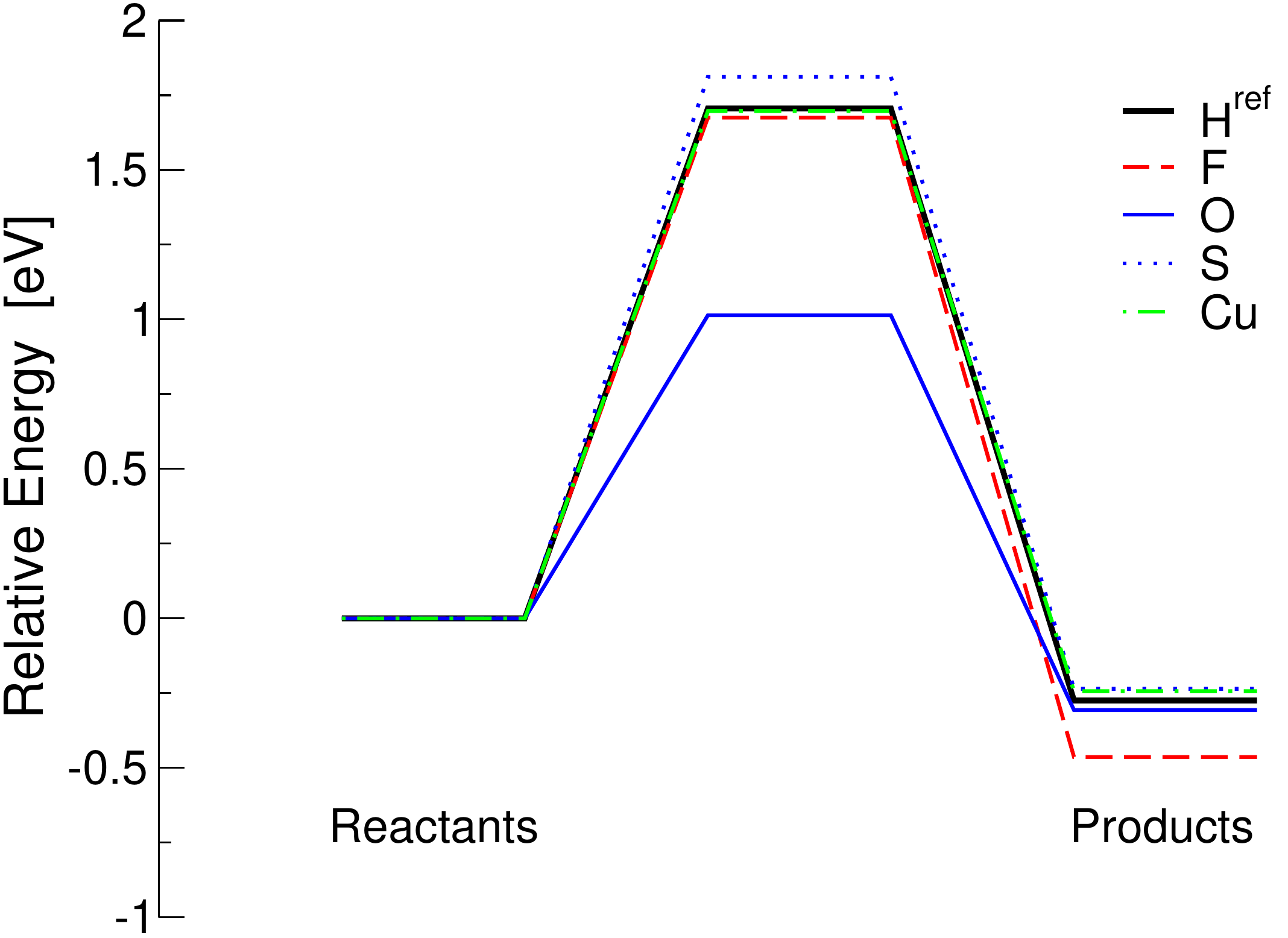}
\caption{\label{fig:barrier_gas}Illustration of the kinetics and thermodynamics in the desorption of H$_2$ from \ABx\ with $\mathcal{X}=$ H, F, O, S, and Cu in the gas phase.}
\end{figure}

In the gas phase, we performed NEB calculations to find the amount of
energy required to remove two H atoms from \AB\ to produce NH$_2$BH$_2$
and H$_2$ as well as the corresponding reaction in substituted \ABx ;
results are collected in Table \ref{tab:NEB_gas} and Fig.~\ref{fig:barrier_gas}. In the standard AB
molecule, the 1.706~eV barrier compares very well with the
37~kcal/mol~$\approx$~1.6~eV barrier found previously at the
CCSD(T)/complete basis set (CBS) level of
theory,\cite{Nguyen_2007:computational_study} as well as other DFT
methods.\cite{Shevlin_2011:dehydrogenation_mechanisms} 
The most apparent change here is the aforementioned barrier decrease for the O
substitution. Whereas other substitutions alter the barrier by no more
than 0.1~eV, O causes a drop of about 0.7~eV. Notably, the resulting
barrier is almost the same as the binding energy for B1 and B2 hydrogen
in the molecule (see Table~\ref{tab:gas_bind}).

In addition to the kinetic reaction barriers, Fig.~\ref{fig:barrier_gas} also reveals
the thermodynamics of the hydrogen release reaction. Our value for $\Delta H$ in pure AB is in very good agreement with numbers in the literature ($-0.28$~eV versus $-0.25$,\cite{Miranda_2007:ab_initio} $-0.29$,\cite{Dixon_2005:thermodynamic_properties} and $-0.30$\cite{Matus_2009:fundamental_thermochemical}). We find that the reaction is slightly exothermic in all cases and most substituents result in an energy release of approximately 0.25 to 0.30~eV per H$_2$ and AB molecule; only the F substitution is slightly more exothermic at 0.46~eV.

While decreasing the H$_2$ desorption barrier for an
isolated molecule is a good indication of improved desorption
properties, a better indication is how the material performs in the
solid, which we discuss next.

\section{H$_2$ desorption in the bulk}\label{sec:bulk}

\begin{figure*}
\includegraphics[width=0.29\textwidth]{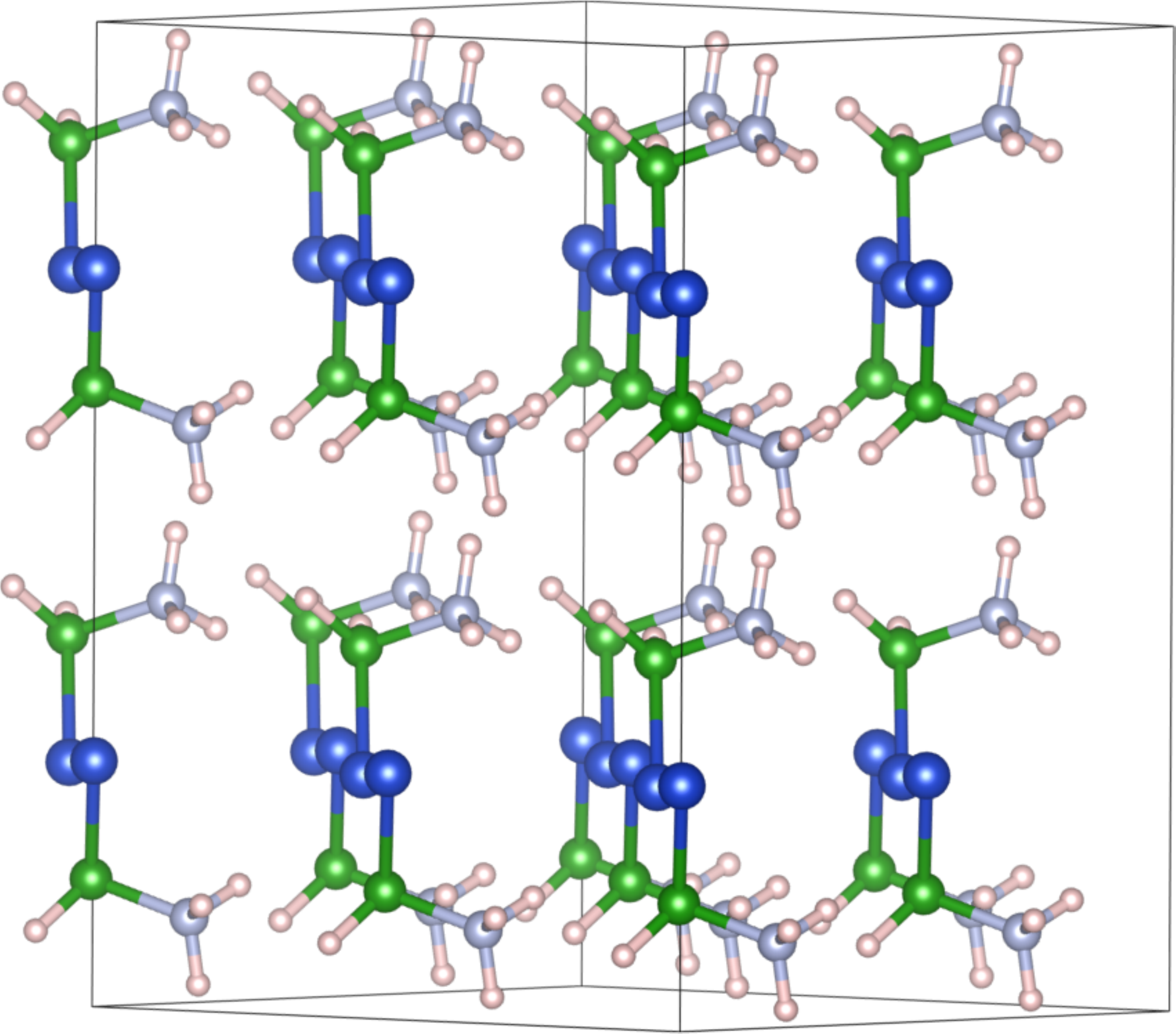}\hfill
\includegraphics[width=0.33\textwidth]{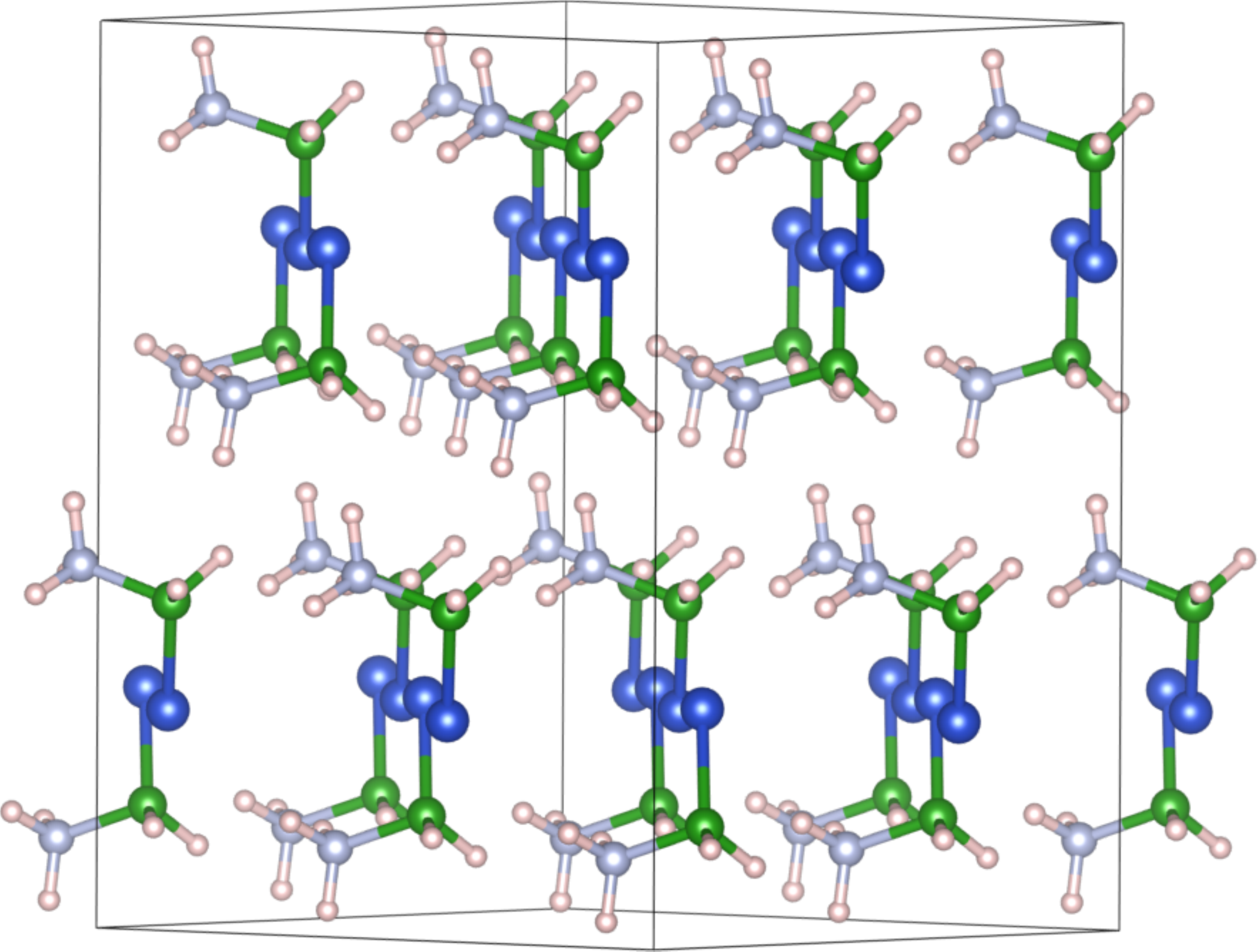}\hfill
\includegraphics[width=0.28\textwidth]{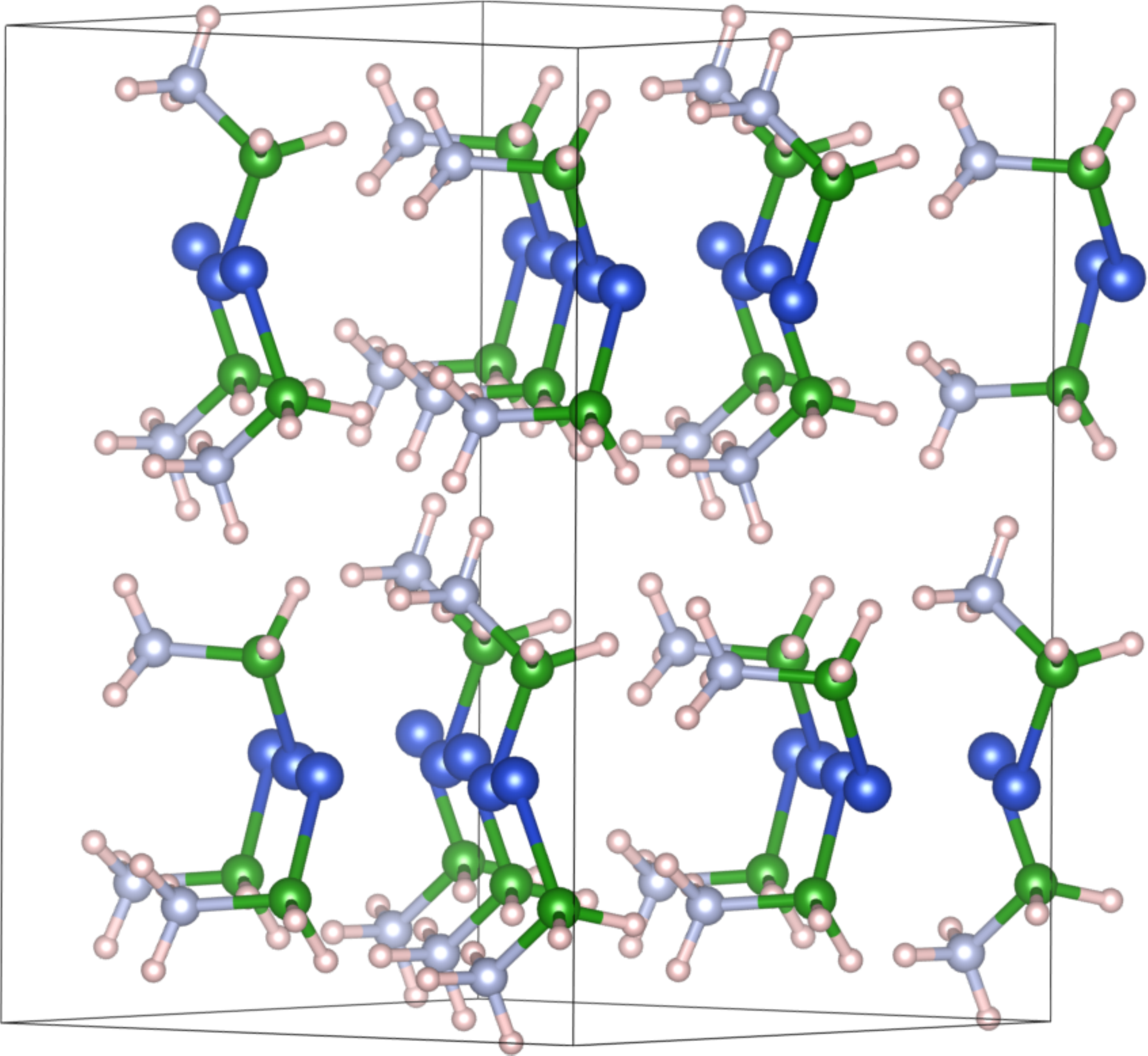}\\
\hspace*{\fill}a)\hfill\hfill b)\hfill\hfill c)\hfill\mbox{}
\caption{\label{fig:Cu-subSolid} Structures found for the Cu-substituted
material and used to calculate the H$_2$ desorption barriers given in
Table~\ref{tab:NEB_solid}. Structure (a) is a supercell of the structure
resulting from the randomized structure search. (b) and (c) are stable
modifications only possible in a larger unit cell. All three structures
are within $\sim$50~meV per molecule energetically and could potentially
coexist in the solid, while also maintaining the geometry of
NH$_3$BH$_2$. Blue atoms are Cu, green are B, purple are N, and pink are
H.}
\end{figure*}

Other studies have focused on hydrogen desorption of monomers and dimers
in the gas phase of \AB.\cite{Nguyen_2007:computational_study,Shevlin_2011:dehydrogenation_mechanisms} However, with a
plane-wave implementation of DFT we have the ability to find transition
states for hydrogen desorption in the solid phase of \AB, which is what
we ultimately want to know, accounting for effects of the entire
environment on the reaction.

\subsection{Structure search}\label{sec:search}

None of the proposed substituted structures \ABx\ have been synthesized
as of yet.  Therefore, we must find a prospective ground-state structure
to simulate the hydrogen desorption reaction. To that end, we first
performed an initial directed structure search for F, O, Cu, and S
substitutions by modifying the experimental low-temperature ground state
of AB. The F, Cu, and S substitutions resulted in (meta)stable candidate
structures that we could use to estimate desorption barriers, with no
imaginary frequencies in their vibrational spectra. We will include
results for these structures in the following tables, but show them with
asterisks in order to indicate that theses structures are derived from
\AB. The O substitution produced a structure where the \ABx molecule is not computationally
stable, with H$_2$ desorption occurring spontaneously during structural
optimization. From this result, we can say that an O substitution seems
to make some of the H atoms in the molecule easier to access (as already
evident from Table~\ref{tab:gas_bind}), however the effect is too
strong. This is the reason S was included in the list of possible
substituents; we want to lower the barrier to H$_2$ desorption, and
ideally the effect of S would be similar to, but less pronounced than that
of O.

In addition, we have also performed a randomized structure search using
the USPEX package. A first search was done with plain AB, verifying that
the method found the experimental low-temperature ground-state
structure.\cite{Bowden_2007:room-temperature_structure} Further searches were done for F, O,
Cu, and S substitutions. The primary objective of these searches was to
find structures that maintained the basic geometry of the original \AB\
molecule, for best comparison with un-doped AB as they are more likely
to retain the hydrogen desorption properties of AB. Searches with F, O,
and S substitutions found that the most energetically favorable
structures significantly deviate from the desired molecular geometry.
For this reason, we only show the Cu structures used (see
Fig.~\ref{fig:Cu-subSolid}), which clearly maintain the NH$_3$BH$_2$
geometry we want to preserve. 

Table \ref{tab:bader_solid} presents a Bader charge analysis for the solid structures, similar to that done on the gas phase structures in Table~\ref{tab:bader}. Most notably, Cu atoms become electron donors in the solid rather than acceptors.

\begin{table}
\caption{\label{tab:bader_solid} Bader charge analysis for solid-phase
structures of \ABx. Numbers given demonstrate the change in charge after
the substitution is made at position B3 and reflect the number of
electrons gained relative to the neutral atom. An asterisk indicates
that results are given for a structure derived from the pure \AB\
structure, which may not be the absolute ground state.}
\begin{tabular*}{\columnwidth}{@{} l @{\extracolsep{\fill}} ccccccc r @{}}\hline
$\mathcal{X}$  & N1 & N2 & N3 & N & B & B1 & B2 & B3$^\text{sub}$\\\hline
H$^\text{ref}$ & $-1.00$ & $-1.00$ & $-1.00$ & 3.07 & $-3.00$ & 0.98 & 0.98 & $ 0.97$\\
         F$^*$ & $-1.00$ & $-1.00$ & $-1.00$ & 3.07 & $-3.00$ & 1.03 & 0.97 & $0.93$ \\
         S$^*$ & $-1.00$ & $-1.00$ & $-1.00$ & 3.11 & $-3.00$ & 1.02 & 1.05 & 0.83 \\
            Cu & $-1.00$ & $-1.00$ & $-1.00$ & 3.11 & $-2.04$ & 1.05 & 0.99 & $-0.11$\\
\end{tabular*}
\end{table}

\subsection{H$_2$ release barriers in the bulk}\label{sec:bulkbarriers}

\begin{figure}
\includegraphics[width=\columnwidth]{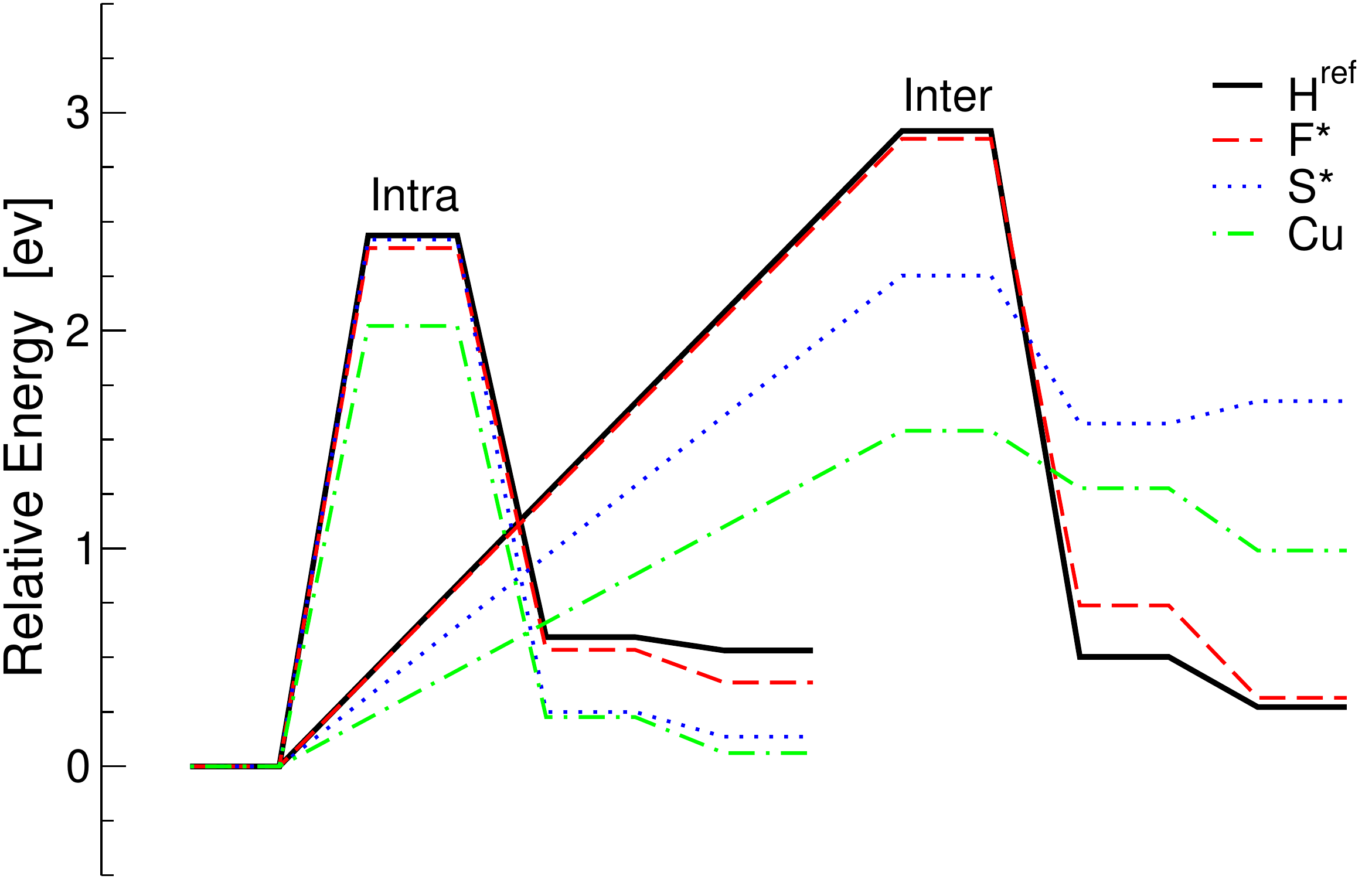}
\caption{\label{fig:barrier_bulk}Illustration of the kinetics and thermodynamics in the desorption of H$_2$ from \ABx\ with $\mathcal{X}=$ H, F, O, S, and Cu in the solid phase via intra- and
inter-molecular pathways, resulting in one molecule of H$_2$ in the supercell, with the two plateaus at the end of the reaction indicating the relative energy of the system before and after the H$_2$ molecule diffuses out of the material.}
\end{figure}

While the precise desorption pathways for AB have long remained
mysterious, the simplest form of this reaction could in general occur in two ways: either two
H atoms come together from opposite ends of an AB molecule to form H$_2$
in an intramolecular reaction, or H$_2$ could form from H atoms in
neighboring molecules in an intermolecular reaction. One recent
theoretical study\cite{Zhong_2012:first-principles_investigation} used
enthalpy of formation calculations to conclude that intermolecular
desorption pathways are more energetically favorable, but did not
consider kinetic barriers. Another
study\cite{Liang_2012:first-principles_study} performed molecular
dynamics simulations, observed an intramolecular desorption event, and
found an energy barrier for the process, but did not consider
intermolecular processes. A more recent experiment found
that AB molecules polymerize in a ``head-to-tail'' manner
through a dehydrocyclization
reaction,\cite{Kobayashi_2014:mechanism_solid-state} concluding
that initial H$_2$ desorption occurs through intermolecular
interactions.
We consider
both possibilities. We have performed NEB calculations to
find the energy barrier to both intra- and inter-molecular forms of
H$_2$ desorption. These calculations were performed in a
$2\times2\times2$ supercell in order to minimize inaccuracies from
mirror-image effects. We have considered many possible combinations of H atoms and pathways for
each reaction and the lowest barriers found in each category are shown
in Table~\ref{tab:NEB_solid} and Fig.~\ref{fig:barrier_bulk}. The two end-points shown in Fig.~\ref{fig:barrier_bulk} represent the relative energetics of the system before and after the H$_2$ molecule diffuses out of the material. There could be a small barrier for diffusion, which we did not consider.

\begin{table}
\caption{\label{tab:NEB_solid} Hydrogen desorption barriers in the solid
phase of \ABx\ resulting in one interstitial molecular H$_2$ in the
solid. An asterisk indicates that results are given for a structure
derived from the pure \AB\ structure, which may not be the absolute
ground state.}
\begin{tabular*}{\columnwidth}{@{} l @{\extracolsep{\fill}} c r @{}}\hline
& \multicolumn{2}{c}{bulk H$_2$ desorption barrier [eV]}\\
$\mathcal{X}$ & intra-molecular & inter-molecular\\\hline
H$^\text{ref}$  & 2.436   &  2.916 \\
         F$^*$  & 2.378   &  2.880 \\
         S$^*$  & 2.418   &  2.252 \\
             Cu & 2.021   &  1.540 \\
\end{tabular*}
\end{table}

From our transition-state searches in pure AB, we find support for the conclusion\cite{Zhong_2012:first-principles_investigation} that the inter-molecular desorption process is more thermodynamically favorable, but
we also find significantly lower kinetic barriers (shown in Table~\ref{tab:NEB_solid}) for the intra-molecular reaction, indicating that the intra-molecular process is more likely to occur than the inter-molecular process. We further find that these barriers are larger than those found for the gas-phase molecules, and that this holds true for each substituent except Cu.
The additional height of these barriers is due
to disruption of the material's dihydrogen bond network, which is much
more rigid in the low-temperature phase we are using for our
simulations. In a previous
study,\cite{Welchman_2014:positional_disorder} we found that significant
barriers to disrupting this network (by rotating molecules, for
instance) do not exist in the room-temperature phase of pure AB,
where the desorption reaction occurs experimentally.

However, the most striking result in Table~\ref{tab:NEB_solid} is that the kinetic barrier for inter-molecular H$_2$ desorption in the Cu-substituted material is significantly lower than that of standard AB, even lower than that of an isolated molecule. Figure~\ref{fig:barrier_bulk} shows that the reaction is more endothermic than the other desorption processes studied, but the kinetic barrier is about 37\% lower than that found for pure AB. Even allowing for a more complex decomposition mechanism in pure AB, we thus conclude that NH$_3$BH$_2$Cu likely has a significantly decreased H$_2$ desorption temperature and is a very promising hydrogen storage material that could potentially fulfill the required DOE targets\cite{DOE_Targets_Onboard_2009} if it can be synthesized.

\section{Conclusions}\label{sec:conclusion}

We examine the effects of substituting C, N, O, F, S, and Cu for one
protic H atom in the BH$_3$ group of AB by simulating H$_2$ desorption
barriers in both isolated molecules and solid structures.
While some of the substituents result in structures that were
not interesting to us or were undesirable---for example, O
substitution creates a stable molecule, but shows spontaneous
H$_2$ desorption when put together to form a solid---we found that 
the Cu substitution is very promising. It lowers the hydrogen desorption
barrier significantly compared to pure AB,
potentially improving on an already attractive hydrogen storage
material. We thus encourage the synthesis of NH$_3$BH$_2$Cu so that the
possibility can be tested.

This work was supported in full by NSF Grant No. DMR-1145968.





\bibliography{refs} 

\providecommand*{\mcitethebibliography}{\thebibliography}
\csname @ifundefined\endcsname{endmcitethebibliography}
{\let\endmcitethebibliography\endthebibliography}{}
\begin{mcitethebibliography}{39}
\providecommand*{\natexlab}[1]{#1}
\providecommand*{\mciteSetBstSublistMode}[1]{}
\providecommand*{\mciteSetBstMaxWidthForm}[2]{}
\providecommand*{\mciteBstWouldAddEndPuncttrue}
  {\def\EndOfBibitem{\unskip.}}
\providecommand*{\mciteBstWouldAddEndPunctfalse}
  {\let\EndOfBibitem\relax}
\providecommand*{\mciteSetBstMidEndSepPunct}[3]{}
\providecommand*{\mciteSetBstSublistLabelBeginEnd}[3]{}
\providecommand*{\EndOfBibitem}{}
\mciteSetBstSublistMode{f}
\mciteSetBstMaxWidthForm{subitem}
{(\emph{\alph{mcitesubitemcount}})}
\mciteSetBstSublistLabelBeginEnd{\mcitemaxwidthsubitemform\space}
{\relax}{\relax}

\bibitem[Z\"{u}ttel(2004)]{Zuttel_2004:hydrogen_storage}
A.~Z\"{u}ttel, \emph{Naturwissenschaften}, 2004, \textbf{91}, 157--172\relax
\mciteBstWouldAddEndPuncttrue
\mciteSetBstMidEndSepPunct{\mcitedefaultmidpunct}
{\mcitedefaultendpunct}{\mcitedefaultseppunct}\relax
\EndOfBibitem
\bibitem[Dresselhaus \emph{et~al.}(2004)Dresselhaus, Crabtree, Buchanan,
  Mallouk, Mets, Taylor, Jena, DiSalvo, and
  Zawodzinski]{Basic_Research_Needs_2004}
M.~Dresselhaus, G.~Crabtree, M.~Buchanan, T.~Mallouk, L.~Mets, K.~Taylor,
  P.~Jena, F.~DiSalvo and T.~Zawodzinski, \emph{Basic Research Needs for the
  Hydrogen Economy}, Us department of energy technical report, 2004\relax
\mciteBstWouldAddEndPuncttrue
\mciteSetBstMidEndSepPunct{\mcitedefaultmidpunct}
{\mcitedefaultendpunct}{\mcitedefaultseppunct}\relax
\EndOfBibitem
\bibitem[DOE(2009)]{DOE_Targets_Onboard_2009}
\emph{DOE Targets for Onboard Hydrogen Storage Systems for Light-Duty
  Vehicles}, {US} department of energy retrieved from
  \url{http://energy.gov/sites/prod/files/2014/03/f12/targets\_onboard\_hydro\_storage.pdf},
  2009\relax
\mciteBstWouldAddEndPuncttrue
\mciteSetBstMidEndSepPunct{\mcitedefaultmidpunct}
{\mcitedefaultendpunct}{\mcitedefaultseppunct}\relax
\EndOfBibitem
\bibitem[Harrison \emph{et~al.}(2014)Harrison, Welchman, Chabal, and
  Thonhauser]{Harrison_2014:materials_hydrogen}
D.~Harrison, E.~Welchman, Y.~J. Chabal and T.~Thonhauser, in \emph{The Handbook
  of Clean Energy Systems}, ed. J.~Yan, Wiley, Hoboken, NJ, 2014, vol. 5:
  Energy Storage\relax
\mciteBstWouldAddEndPuncttrue
\mciteSetBstMidEndSepPunct{\mcitedefaultmidpunct}
{\mcitedefaultendpunct}{\mcitedefaultseppunct}\relax
\EndOfBibitem
\bibitem[Graetz(2009)]{Graetz_2009:new_approaches}
J.~Graetz, \emph{Chem. Soc. Rev.}, 2009, \textbf{38}, 73--82\relax
\mciteBstWouldAddEndPuncttrue
\mciteSetBstMidEndSepPunct{\mcitedefaultmidpunct}
{\mcitedefaultendpunct}{\mcitedefaultseppunct}\relax
\EndOfBibitem
\bibitem[Xiong \emph{et~al.}(2008)Xiong, Yong, Wu, Chen, Shaw, Karkamkar,
  Autrey, Jones, Johnson, Edwards, and
  David]{Xiong_2008:high-capacity_hydrogen}
Z.~Xiong, C.~K. Yong, G.~Wu, P.~Chen, W.~J. Shaw, A.~Karkamkar, T.~Autrey,
  M.~O. Jones, S.~R. Johnson, P.~P. Edwards and W.~I.~F. David, \emph{Nat.
  Mater.}, 2008, \textbf{7}, 138--141\relax
\mciteBstWouldAddEndPuncttrue
\mciteSetBstMidEndSepPunct{\mcitedefaultmidpunct}
{\mcitedefaultendpunct}{\mcitedefaultseppunct}\relax
\EndOfBibitem
\bibitem[Chua \emph{et~al.}(2011)Chua, Chen, Wu, and
  Xiong]{Chua_2011:development_amidoboranes}
Y.~S. Chua, P.~Chen, G.~Wu and Z.~Xiong, \emph{Chem. Commun.}, 2011,
  \textbf{47}, 5116--5129\relax
\mciteBstWouldAddEndPuncttrue
\mciteSetBstMidEndSepPunct{\mcitedefaultmidpunct}
{\mcitedefaultendpunct}{\mcitedefaultseppunct}\relax
\EndOfBibitem
\bibitem[Swinnen \emph{et~al.}(2010)Swinnen, Nguyen, and
  Nguyen]{Swinnen_2010:potential_hydrogen}
S.~Swinnen, V.~S. Nguyen and M.~T. Nguyen, \emph{Chem. Phys. Lett.}, 2010,
  \textbf{489}, 148--153\relax
\mciteBstWouldAddEndPuncttrue
\mciteSetBstMidEndSepPunct{\mcitedefaultmidpunct}
{\mcitedefaultendpunct}{\mcitedefaultseppunct}\relax
\EndOfBibitem
\bibitem[Hamilton \emph{et~al.}(2009)Hamilton, Baker, Staubitz, and
  Manners]{Hamilton_2009:b-n_compounds}
C.~W. Hamilton, R.~T. Baker, A.~Staubitz and I.~Manners, \emph{Chem. Soc.
  Rev.}, 2009, \textbf{38}, 279--293\relax
\mciteBstWouldAddEndPuncttrue
\mciteSetBstMidEndSepPunct{\mcitedefaultmidpunct}
{\mcitedefaultendpunct}{\mcitedefaultseppunct}\relax
\EndOfBibitem
\bibitem[Heldebrant \emph{et~al.}(2008)Heldebrant, Karkamkar, Hess, Bowden,
  Rassat, Zheng, Rappe, and Autrey]{Heldebrant_2008:effects_chemical}
D.~J. Heldebrant, A.~Karkamkar, N.~J. Hess, M.~E. Bowden, S.~Rassat, F.~Zheng,
  K.~Rappe and T.~Autrey, \emph{Chem. Mater.}, 2008, \textbf{20},
  5332--5336\relax
\mciteBstWouldAddEndPuncttrue
\mciteSetBstMidEndSepPunct{\mcitedefaultmidpunct}
{\mcitedefaultendpunct}{\mcitedefaultseppunct}\relax
\EndOfBibitem
\bibitem[Marder(2007)]{Marder_2007:will_we}
T.~B. Marder, \emph{Angew. Chemie Int. Ed.}, 2007, \textbf{46},
  8116--8118\relax
\mciteBstWouldAddEndPuncttrue
\mciteSetBstMidEndSepPunct{\mcitedefaultmidpunct}
{\mcitedefaultendpunct}{\mcitedefaultseppunct}\relax
\EndOfBibitem
\bibitem[Stowe \emph{et~al.}(2007)Stowe, Shaw, Linehan, Schmid, and
  Autrey]{Stowe_2007:situ_solid}
A.~C. Stowe, W.~J. Shaw, J.~C. Linehan, B.~Schmid and T.~Autrey, \emph{Phys.
  Chem. Chem. Phys.}, 2007, \textbf{9}, 1831--1836\relax
\mciteBstWouldAddEndPuncttrue
\mciteSetBstMidEndSepPunct{\mcitedefaultmidpunct}
{\mcitedefaultendpunct}{\mcitedefaultseppunct}\relax
\EndOfBibitem
\bibitem[Kobayashi \emph{et~al.}(2014)Kobayashi, Gupta, Caporini, Pecharsky,
  and Pruski]{Kobayashi_2014:mechanism_solid-state}
T.~Kobayashi, S.~Gupta, M.~A. Caporini, V.~K. Pecharsky and M.~Pruski, \emph{J.
  Phys. Chem. C}, 2014, \textbf{118}, 19548--19555\relax
\mciteBstWouldAddEndPuncttrue
\mciteSetBstMidEndSepPunct{\mcitedefaultmidpunct}
{\mcitedefaultendpunct}{\mcitedefaultseppunct}\relax
\EndOfBibitem
\bibitem[Yang \emph{et~al.}(2010)Yang, Sudik, Wolverton, and
  Siegel]{Yang_2010:high_capacity}
J.~Yang, A.~Sudik, C.~Wolverton and D.~J. Siegel, \emph{Chem. Soc. Rev.}, 2010,
  \textbf{39}, 656--675\relax
\mciteBstWouldAddEndPuncttrue
\mciteSetBstMidEndSepPunct{\mcitedefaultmidpunct}
{\mcitedefaultendpunct}{\mcitedefaultseppunct}\relax
\EndOfBibitem
\bibitem[Allred(1961)]{Allred_1961:electronegativity_values}
A.~L. Allred, \emph{J. Inorg. Nucl. Chem.}, 1961, \textbf{17}, 215--221\relax
\mciteBstWouldAddEndPuncttrue
\mciteSetBstMidEndSepPunct{\mcitedefaultmidpunct}
{\mcitedefaultendpunct}{\mcitedefaultseppunct}\relax
\EndOfBibitem
\bibitem[Shevlin \emph{et~al.}(2011)Shevlin, Kerkeni, and
  Guo]{Shevlin_2011:dehydrogenation_mechanisms}
S.~A. Shevlin, B.~Kerkeni and Z.~X. Guo, \emph{Phys. Chem. Chem. Phys.}, 2011,
  \textbf{13}, 7649--7659\relax
\mciteBstWouldAddEndPuncttrue
\mciteSetBstMidEndSepPunct{\mcitedefaultmidpunct}
{\mcitedefaultendpunct}{\mcitedefaultseppunct}\relax
\EndOfBibitem
\bibitem[Zhong \emph{et~al.}(2012)Zhong, Song, Huang, Xia, and
  Wen]{Zhong_2012:first-principles_investigation}
B.~Zhong, L.~Song, X.~X. Huang, L.~Xia and G.~Wen, \emph{Physica Scripta},
  2012, \textbf{86}, 015606\relax
\mciteBstWouldAddEndPuncttrue
\mciteSetBstMidEndSepPunct{\mcitedefaultmidpunct}
{\mcitedefaultendpunct}{\mcitedefaultseppunct}\relax
\EndOfBibitem
\bibitem[Miranda and Ceder(2007)]{Miranda_2007:ab_initio}
C.~R. Miranda and G.~Ceder, \emph{J. Chem. Phys.}, 2007, \textbf{126},
  184703\relax
\mciteBstWouldAddEndPuncttrue
\mciteSetBstMidEndSepPunct{\mcitedefaultmidpunct}
{\mcitedefaultendpunct}{\mcitedefaultseppunct}\relax
\EndOfBibitem
\bibitem[Nguyen \emph{et~al.}(2007)Nguyen, Matus, Grant, Nguyen, and
  Dixon]{Nguyen_2007:computational_study}
V.~S. Nguyen, M.~H. Matus, D.~J. Grant, M.~T. Nguyen and D.~A. Dixon, \emph{J.
  Phys. Chem. A}, 2007, \textbf{111}, 8844--8856\relax
\mciteBstWouldAddEndPuncttrue
\mciteSetBstMidEndSepPunct{\mcitedefaultmidpunct}
{\mcitedefaultendpunct}{\mcitedefaultseppunct}\relax
\EndOfBibitem
\bibitem[Wolstenholme \emph{et~al.}(2012)Wolstenholme, Traboulsee, Hua,
  Calhoun, and McGrady]{Wolstenholme_2012:thermal_desorption}
D.~J. Wolstenholme, K.~T. Traboulsee, Y.~Hua, L.~A. Calhoun and G.~S. McGrady,
  \emph{Chem. Commun.}, 2012, \textbf{48}, 2597--2599\relax
\mciteBstWouldAddEndPuncttrue
\mciteSetBstMidEndSepPunct{\mcitedefaultmidpunct}
{\mcitedefaultendpunct}{\mcitedefaultseppunct}\relax
\EndOfBibitem
\bibitem[Choi \emph{et~al.}(2014)Choi, R{\"o}nnebro, Rassat, Karkamkar, Maupin,
  Holladay, Simmons, and Brooks]{Choi_2014:kinetics_study}
Y.~J. Choi, E.~C.~E. R{\"o}nnebro, S.~Rassat, A.~Karkamkar, G.~Maupin,
  J.~Holladay, K.~Simmons and K.~Brooks, \emph{Phys. Chem. Chem. Phys.}, 2014,
  \textbf{16}, 7959--7968\relax
\mciteBstWouldAddEndPuncttrue
\mciteSetBstMidEndSepPunct{\mcitedefaultmidpunct}
{\mcitedefaultendpunct}{\mcitedefaultseppunct}\relax
\EndOfBibitem
\bibitem[Kresse and Furthm\"{u}ller(1996)]{Kresse_1996:efficient_iterative}
G.~Kresse and J.~Furthm\"{u}ller, \emph{Phys. Rev. B}, 1996, \textbf{54},
  11169--11186\relax
\mciteBstWouldAddEndPuncttrue
\mciteSetBstMidEndSepPunct{\mcitedefaultmidpunct}
{\mcitedefaultendpunct}{\mcitedefaultseppunct}\relax
\EndOfBibitem
\bibitem[Kresse and Joubert(1999)]{Kresse_1999:ultrasoft_pseudopotentials}
G.~Kresse and D.~Joubert, \emph{Phys. Rev. B}, 1999, \textbf{59},
  1758--1775\relax
\mciteBstWouldAddEndPuncttrue
\mciteSetBstMidEndSepPunct{\mcitedefaultmidpunct}
{\mcitedefaultendpunct}{\mcitedefaultseppunct}\relax
\EndOfBibitem
\bibitem[Chen \emph{et~al.}(2010)Chen, Couvy, Liu, Drozd, Daemen, Zhao, and
  Kao]{Chen_2010:situ_x-ray}
J.~Chen, H.~Couvy, H.~Liu, V.~Drozd, L.~L. Daemen, Y.~Zhao and C.-C. Kao,
  \emph{Int. J. Hydrogen Energy}, 2010, \textbf{35}, 11064--11070\relax
\mciteBstWouldAddEndPuncttrue
\mciteSetBstMidEndSepPunct{\mcitedefaultmidpunct}
{\mcitedefaultendpunct}{\mcitedefaultseppunct}\relax
\EndOfBibitem
\bibitem[Lin \emph{et~al.}(2008)Lin, Mao, Drozd, Chen, and
  Daemen]{Lin_2008:raman_spectroscopy}
Y.~Lin, W.~L. Mao, V.~Drozd, J.~Chen and L.~L. Daemen, \emph{J. Chem. Phys.},
  2008, \textbf{129}, 234509\relax
\mciteBstWouldAddEndPuncttrue
\mciteSetBstMidEndSepPunct{\mcitedefaultmidpunct}
{\mcitedefaultendpunct}{\mcitedefaultseppunct}\relax
\EndOfBibitem
\bibitem[Dion \emph{et~al.}(2004)Dion, Rydberg, Schr\"{o}der, Langreth, and
  Lundqvist]{Dion_2004:van_waals}
M.~Dion, H.~Rydberg, E.~Schr\"{o}der, D.~C. Langreth and B.~I. Lundqvist,
  \emph{Phys. Rev. Lett.}, 2004, \textbf{92}, 246401\relax
\mciteBstWouldAddEndPuncttrue
\mciteSetBstMidEndSepPunct{\mcitedefaultmidpunct}
{\mcitedefaultendpunct}{\mcitedefaultseppunct}\relax
\EndOfBibitem
\bibitem[Thonhauser \emph{et~al.}(2007)Thonhauser, Cooper, Li, Puzder,
  Hyldgaard, and Langreth]{Thonhauser_2007:van_waals}
T.~Thonhauser, V.~R. Cooper, S.~Li, A.~Puzder, P.~Hyldgaard and D.~C. Langreth,
  \emph{Phys. Rev. B}, 2007, \textbf{76}, 125112\relax
\mciteBstWouldAddEndPuncttrue
\mciteSetBstMidEndSepPunct{\mcitedefaultmidpunct}
{\mcitedefaultendpunct}{\mcitedefaultseppunct}\relax
\EndOfBibitem
\bibitem[Langreth \emph{et~al.}(2009)Langreth, Lundqvist, Chakarova-K\"{a}ck,
  Cooper, Dion, Hyldgaard, Kelkkanen, Kleis, Kong, Li, Moses, Murray, Puzder,
  Rydberg, Schr\"{o}der, and Thonhauser]{Langreth_2009:density_functional}
D.~C. Langreth, B.~I. Lundqvist, S.~D. Chakarova-K\"{a}ck, V.~R. Cooper,
  M.~Dion, P.~Hyldgaard, A.~Kelkkanen, J.~Kleis, L.~Kong, S.~Li, P.~G. Moses,
  E.~D. Murray, A.~Puzder, H.~Rydberg, E.~Schr\"{o}der and T.~Thonhauser,
  \emph{J. Phys. Condens. Matter}, 2009, \textbf{21}, 084203\relax
\mciteBstWouldAddEndPuncttrue
\mciteSetBstMidEndSepPunct{\mcitedefaultmidpunct}
{\mcitedefaultendpunct}{\mcitedefaultseppunct}\relax
\EndOfBibitem
\bibitem[Berland \emph{et~al.}(2015)Berland, Cooper, Lee, Schr\"oder,
  Thonhauser, Hyldgaard, and Lundqvist]{Berland_2015:van_waals}
K.~Berland, V.~R. Cooper, K.~Lee, E.~Schr\"oder, T.~Thonhauser, P.~Hyldgaard
  and B.~I. Lundqvist, \emph{Rep. Prog. Phys.}, 2015\relax
\mciteBstWouldAddEndPuncttrue
\mciteSetBstMidEndSepPunct{\mcitedefaultmidpunct}
{\mcitedefaultendpunct}{\mcitedefaultseppunct}\relax
\EndOfBibitem
\bibitem[Oganov and Glass(2006)]{Oganov_2006:crystal_structure}
A.~R. Oganov and C.~W. Glass, \emph{J. Chem. Phys.}, 2006, \textbf{124},
  244704\relax
\mciteBstWouldAddEndPuncttrue
\mciteSetBstMidEndSepPunct{\mcitedefaultmidpunct}
{\mcitedefaultendpunct}{\mcitedefaultseppunct}\relax
\EndOfBibitem
\bibitem[Lyakhov \emph{et~al.}(2013)Lyakhov, Oganov, Stokes, and
  Zhu]{Lyakhov_2013:new_developments}
A.~O. Lyakhov, A.~R. Oganov, H.~T. Stokes and Q.~Zhu, \emph{Comput. Phys.
  Commun.}, 2013, \textbf{184}, 1172--1182\relax
\mciteBstWouldAddEndPuncttrue
\mciteSetBstMidEndSepPunct{\mcitedefaultmidpunct}
{\mcitedefaultendpunct}{\mcitedefaultseppunct}\relax
\EndOfBibitem
\bibitem[Zhu \emph{et~al.}(2012)Zhu, Oganov, Glass, and
  Stokes]{Zhu_2012:constrained_evolutionary}
Q.~Zhu, A.~R. Oganov, C.~W. Glass and H.~T. Stokes, \emph{Acta Crystallogr.
  Sect. B Struct. Sci.}, 2012, \textbf{B68}, 215--226\relax
\mciteBstWouldAddEndPuncttrue
\mciteSetBstMidEndSepPunct{\mcitedefaultmidpunct}
{\mcitedefaultendpunct}{\mcitedefaultseppunct}\relax
\EndOfBibitem
\bibitem[Henkelman \emph{et~al.}(2000)Henkelman, Uberuaga, and
  J\'onsson]{Henkelman_2000:climbing_image}
G.~Henkelman, B.~P. Uberuaga and H.~J\'onsson, \emph{J. Chem. Phys.}, 2000,
  \textbf{113}, 9901--9904\relax
\mciteBstWouldAddEndPuncttrue
\mciteSetBstMidEndSepPunct{\mcitedefaultmidpunct}
{\mcitedefaultendpunct}{\mcitedefaultseppunct}\relax
\EndOfBibitem
\bibitem[Henkelman and J\'onsson(2000)]{Henkelman_2000:improved_tangent}
G.~Henkelman and H.~J\'onsson, \emph{J. Chem. Phys.}, 2000, \textbf{113},
  9978--9985\relax
\mciteBstWouldAddEndPuncttrue
\mciteSetBstMidEndSepPunct{\mcitedefaultmidpunct}
{\mcitedefaultendpunct}{\mcitedefaultseppunct}\relax
\EndOfBibitem
\bibitem[Dixon and Gutowski(2005)]{Dixon_2005:thermodynamic_properties}
D.~A. Dixon and M.~Gutowski, \emph{J. Phys. Chem. A}, 2005, \textbf{109},
  5129--5135\relax
\mciteBstWouldAddEndPuncttrue
\mciteSetBstMidEndSepPunct{\mcitedefaultmidpunct}
{\mcitedefaultendpunct}{\mcitedefaultseppunct}\relax
\EndOfBibitem
\bibitem[Matus \emph{et~al.}(2009)Matus, Grant, Nguyen, and
  Dixon]{Matus_2009:fundamental_thermochemical}
M.~H. Matus, D.~J. Grant, M.~T. Nguyen and D.~A. Dixon, \emph{J. Phys. Chem.
  C}, 2009, \textbf{113}, 16553--16560\relax
\mciteBstWouldAddEndPuncttrue
\mciteSetBstMidEndSepPunct{\mcitedefaultmidpunct}
{\mcitedefaultendpunct}{\mcitedefaultseppunct}\relax
\EndOfBibitem
\bibitem[Bowden \emph{et~al.}(2007)Bowden, Gainsford, and
  Robinson]{Bowden_2007:room-temperature_structure}
M.~E. Bowden, G.~J. Gainsford and W.~T. Robinson, \emph{Aust. J. Chem.}, 2007,
  \textbf{60}, 149--153\relax
\mciteBstWouldAddEndPuncttrue
\mciteSetBstMidEndSepPunct{\mcitedefaultmidpunct}
{\mcitedefaultendpunct}{\mcitedefaultseppunct}\relax
\EndOfBibitem
\bibitem[Liang and Tse(2012)]{Liang_2012:first-principles_study}
Y.~Liang and J.~S. Tse, \emph{J. Phys. Chem. C}, 2012, \textbf{116},
  2146--2152\relax
\mciteBstWouldAddEndPuncttrue
\mciteSetBstMidEndSepPunct{\mcitedefaultmidpunct}
{\mcitedefaultendpunct}{\mcitedefaultseppunct}\relax
\EndOfBibitem
\bibitem[Welchman \emph{et~al.}(2014)Welchman, Giannozzi, and
  Thonhauser]{Welchman_2014:positional_disorder}
E.~Welchman, P.~Giannozzi and T.~Thonhauser, \emph{Phys. Rev. B}, 2014,
  \textbf{89}, 180101(R)\relax
\mciteBstWouldAddEndPuncttrue
\mciteSetBstMidEndSepPunct{\mcitedefaultmidpunct}
{\mcitedefaultendpunct}{\mcitedefaultseppunct}\relax
\EndOfBibitem
\end{mcitethebibliography}
\bibliographystyle{rsc} 

\end{document}